\documentclass[%
 aip,
 amsmath,amssymb,
reprint, onecolumn 
]{revtex4-1}

\usepackage{graphicx}
\usepackage{dcolumn}
\usepackage{bm}

\usepackage[utf8]{inputenc}
\usepackage[T1]{fontenc}
\usepackage{mathptmx}
\usepackage{etoolbox}

\makeatletter
\def\@email#1#2{%
 \endgroup
 \patchcmd{\titleblock@produce}
  {\frontmatter@RRAPformat}
  {\frontmatter@RRAPformat{\produce@RRAP{*#1\href{mailto:#2}{#2}}}\frontmatter@RRAPformat}
  {}{}
}%
\makeatother
\begin{document}

\preprint{AIP/123-QED}

\title[]{Simplified approach to estimate Lorenz number using experimental Seebeck coefficient for non parabolic band}
\author{Ankit Kumar}
 \altaffiliation[]{}
\email{kumar.ankit@students.iiserpune.ac.in}
\affiliation{Department of Physics, Indian Institute of Science Education and Research, Pune, Maharashtra, India, 411008}%

\date{\today}

\begin{abstract}
Reduction of lattice thermal conductivity ($\kappa_L$) is one of the most effective ways of improving thermoelectric properties. However extraction of $\kappa_L$ from the total measured thermal conductivity can be misleading if Lorenz ($L$) number is not estimated correctly. The $\kappa_L$ is obtained using Wiedemann-Franz law which estimates electronic part of thermal conductivity $\kappa_e$ = $L$$\sigma$T where, $\sigma$ and T are electrical conductivity and temperature. The $\kappa_L$ is then estimated as $\kappa_L$ = $\kappa_T$ - $L$$\sigma$T. For the metallic system the Lorenz number has universal value of 2.44 $\times$ 10$^{-8}$ W$\Omega$K$^{-2}$ (degenerate limit), but for no-degenerate semiconductors, the value can deviate significantly for acoustic phonon scattering, the most common scattering mechanism for thermoelectric above room temperatures. Up till now, $L$ is estimated by solving a series of equation derived form Boltzmann transport equations. For the single parabolic band (SPB) an equation was proposed to estimate $L$ directly from the experimental Seebeck coefficient. However using SPB model will lead to overestimation of $L$ in case of low band gap semiconductors which result in underestimation of $\kappa_L$ sometimes even negative $\kappa_L$. In this letter we propose a simpler equation to estimate $L$ for a non parabolic band. Experimental Seebeck coefficient, band gap($E_g$), and Temperature ($T$) are the main inputs in the equation which nearly eliminates the need of solving multiple Fermi integrals besides giving accurate values of $L$. 
\end{abstract}

\date{\today}

\draft 
\maketitle 


Direct conversion of heat into electricity is done using thermoelectric materials. The conversion efficiency is dependent on a dimensionless quantity called figure of merit (zT) = $S^2\sigma T$/($\kappa_e$ + $\kappa_L$), where $S$, $\sigma$, $T$, $\kappa_e$ and $\kappa_L$ are Seebeck coefficient, electrical conductivity, absolute temperature, electronic thermal conductivity, and lattice thermal conductivity respectively. Typical measurement of thermal conductivity gives the total value which is $\kappa_T$ = $\kappa_e$ + $\kappa_L$. Using measured $\sigma$ and the Wiedemann-Franz law, $\kappa_e$ is estimated as: $\kappa_e$ = L$\sigma$T, where $L$ is the Lorenz number~\cite{jones1985theoretical}. Once $\kappa_e$ is known, $\kappa_L$ can be determined by removing the electronic contribution from total thermal conductivity, $\kappa_T$ - L$\sigma$T = $\kappa_L$. In the case of a bipolar system, $\kappa_T$  - L$\sigma$T = $\kappa_L$ + $\kappa_{bipolar}$ and hence the knowledge of $\kappa_{bipolar}$ becomes necessary for the exact estimation of $\kappa_L$. However, in most cases of interest, either $\kappa_{bipolar}$ is small or is not there at all.

In order to get a high zT value, $\kappa_T$ has to be reduced but $\sigma$ has to be increased. As $\kappa_e$ and $\sigma$ are directly correlated, reducing $\kappa_L$ is an effective way to increase zT.~\cite{snyder2008complex} However, estimation of $\kappa_L$ using $L$ can often be misleading. If one uses the value of $L$ used for a metallic system for semiconductors (SPB) where the actual $L$ is less than 2.44 $\times$ 10$^{-8}$ W$\Omega$K$^{-2}$, the estimated $\kappa_L$ comes out to be less than actual value. Similarly for non parabolic bands $L$ is even smaller than that of SBP.  In that case if $L$ is estimated using SPB model it will give $\kappa_L$ to be quite less. These results can be misleading in the scene that the actual $\kappa_L$ is not that low. For example, incautious determination of $L$ in case of lanthanum telluride can even cause $\kappa_L$ to be negative, which is not physical~\cite{may2008thermoelectric}. Therefore, careful evaluation of $L$ is critical in characterizing enhancements in zT due to $\kappa_L$ reduction. 

In case of metals, charge carriers are free-electrons like, where $L$ converges to 2.44 $\times$ 10$^{-8}$ W$\Omega$ K$^{-2}$ (degenerate limit). However, heavily doped semiconductors will have $L$ very close to the degenerate limit. From thermoelectric point of view, most good thermoelectric materials have their carrier densities between lightly doped and heavily doped regions. In such cases the error in $L$ could be as much as 40 \% ~\cite{toberer2012advances}.
Measuring $L$ directly requires high mobility which is beyond attainable above room temperature~\cite{lukas2012experimental}. So far, $L$ is taken as a constant (2.44 10$^{-8}$ W$\Omega$ K$^{-2}$) or estimated using various band models such as the single parabolic band model (SPB) and single Kane band model (SKB) which considers non parabolic nature of bands. 
Although both SPB and SKB models work well to estimate $L$, a transcendental set of equations is needs to be solved for $L$ in terms of S that require a numerical solution. 
By assuming that the carrier relaxation time is limited by acoustic phonon scattering (APS), one of the most relevant scattering mechanisms for thermoelectric materials above room temperature, followings are the equations that needs to be solved for SPB-APS model~\cite{pei2011convergence, wood1988materials}:

\begin{equation}
	S = \frac{k\textsubscript{B} }{e}\left[ \frac{2F\textsubscript{1}(\eta)}{F\textsubscript{0}(\eta)} -\eta \right],
\end{equation}
\begin{equation}
	L = \left(\frac{k\textsubscript{B}}{e}\right)^2\left[\frac{3F\textsubscript{2}(\eta)}{F\textsubscript{0}(\eta)} - \left(\frac{2F\textsubscript{1}(\eta)}{F\textsubscript{0}(\eta)} \right)^2\right]
\end{equation}

\begin{equation}
	\eta = \frac{E\textsubscript{F} }{k\textsubscript{B}T},
\end{equation}
where, $\eta$ is the reduced Fermi energy, $E_F$ the Fermi energy, $F_n$($\eta)$ is the Fermi integral, $k_B$ is the Boltzmann constant , and e is the electronic charge.

\begin{equation}
    F_{k}(\eta) = \int_{0}^{\infty}{\frac{e^k \partial \varepsilon}{1 + exp(E - \eta)}}
\end{equation} 

For SPB model, a much simpler form is proposed by Hyun-Sik Kim et.al which estimated $L$ within 5 \% error with respect to actual value estimated using SPB model\cite{kim2015characterization}. The expression for $L$ as derived by Kim et.al. is given as: 

\begin{equation}
	L = 1.5 + exp \left(-\frac{|S|}{116}\right)
\end{equation}

However, the SPB model with acoustic phonon scattering does not produce correct results in case of low band gap semiconductors where non-parabolic band structure come into picture. PbTe~\cite{pei2011convergence, lalonde2011reevaluation}, PbSe~\cite{wang2012weak}, and PbS~\cite{wang2013high} are a few examples with narrow band gap described by non-parabolic Kane band model. The non-parabolicity parameter is determined as: $\alpha$ = $k_B$T/$E_g$, where $E_g$ is the band gap~\cite{bhandari1985electronic, ravich2013semiconducting}. Estimation of $L$ using SPB model produced value 26 \% less~\cite{kim2015characterization} than actual. With overestimated $L$, $\kappa_e$ will be overestimated too. Therefore when $\kappa_e$ is subtracted from $\kappa_T$ to get  $\kappa_L$, it will be highly underestimated. 
Considering non parabilicity into account provides the correct estimation. The Seebeck coefficient and $L$ for for a non-parabolic band is given as:
\begin{equation}\label{see}
	S = \frac{k\textsubscript{B} }{e}\left[\frac{^0F^1_{-2}(\eta, \alpha)}{^1F^1_{-2}(\eta, \alpha)} -\eta\right]
\end{equation}
\begin{equation}\label{l}
	L = \left(\frac{k_{B}}{e}\right)^2\left[\frac{^2F^1_{-2}(\eta, \alpha)}{^0F^1_{-2}(\eta, \alpha)} - \left(\frac{^1F^1_{-2}(\eta, \alpha)}{^0F^1_{-2}(\eta,\alpha)} \right)^2\right]
\end{equation}
\begin{figure}
	\centering
	\includegraphics[width=1\textwidth]{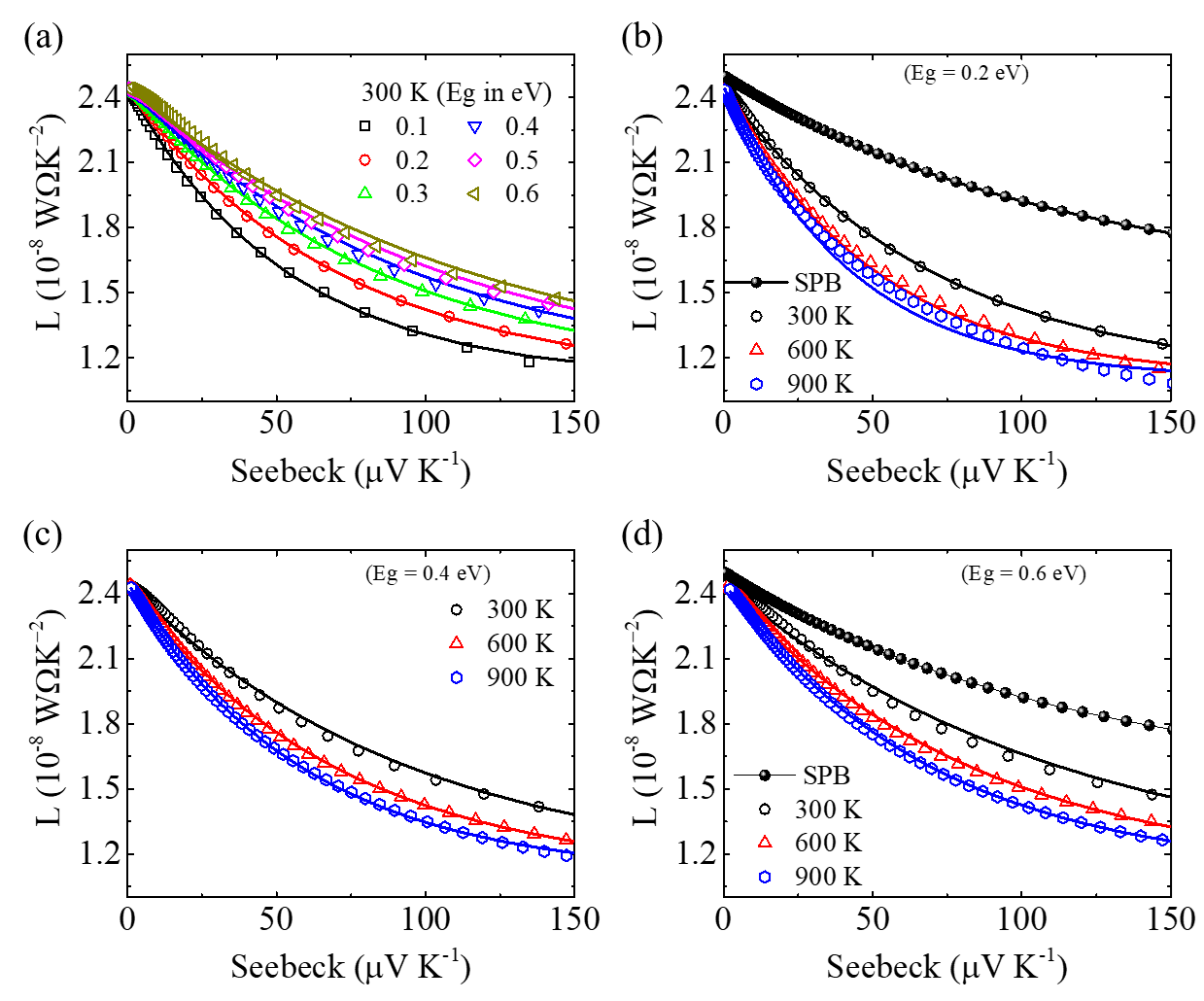}
	\caption {(a) Lorenz number for different $E_g$ at 300 K, and (b, c, d) $L$ for different $E_g$ at different temperatures. Dots are the calculated date using the SKB model and the solid lines are using equation~\ref{new}}
	\label{fit}   
\end{figure}

\begin{figure}
	\centering
	\includegraphics[width=0.5\textwidth]{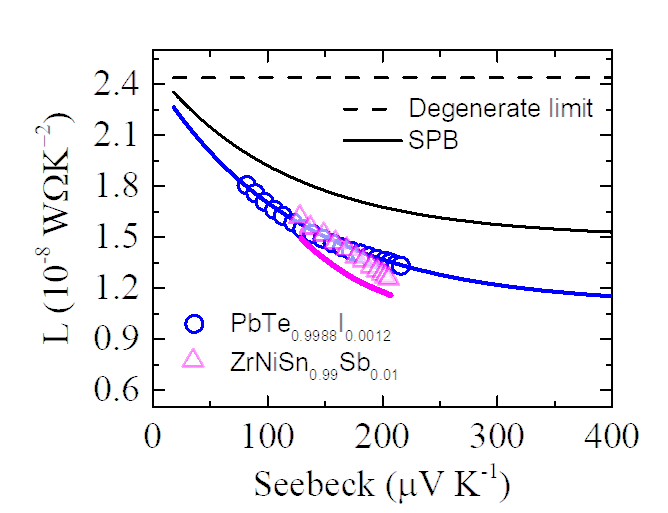}
	\caption {Reported Lorenz number (points) and calculated using equation~\ref{new}(lines). The dotted line represents the degenerate limit whereas the solid black line represents the value of $L$ considering the SPB model.}
	\label{example}   
\end{figure} 

\begin{equation}
    \eta = \frac{E_{F} }{k_{B}T}, ~~~and ~~ \alpha = \frac{k_{B}T}{E_{g}}
\end{equation} 

The function $^nF^m_{k}(\eta,\alpha)$ is the generalized Fermi integral given as:

\begin{equation}
    ^nF^m_{k}(\eta,\alpha) = \int_{0}^{\infty}{-\left(\frac{\partial f}{\partial \varepsilon}\right) \varepsilon\textsuperscript{n}(\varepsilon +\alpha \varepsilon\textsuperscript{2})\textsuperscript{m}(1 + 2\alpha\varepsilon\textsuperscript{2})\textsuperscript{k}\partial \varepsilon}
\end{equation} 

where $\varepsilon$ is the reduced energy $E/k_BT$, and $f$ is the Fermi distribution function. The parameter n, m and k are the indices of the integral whose value depend on the transport property and scattering mechanism.

The SKB model is much more complex than the SPB model as it has additional parameter of non parabolicity ($\alpha$). Considering the exponential form of $L$ for the SPB model, we assumed a similar function of $L$ for SKB with additional parameter of temperature ($T$) and band gap ($E_g$) or in other words $\alpha$. The assumed form for $L$ is given as:
\begin{equation} \label{new}
    L  = A + Bexp \left[\frac{-|S|}{f(E_g)g(T)} \right]
\end{equation}
where $A$ and $B$ are constants. The $T$ and $E_g$ dependence is accounted by function $f$ and $g$. 
In order to find the unknown constants ($A, B$) and functional form ($f, g$), we have generated a data set using equation~\ref{see} and~\ref{l} for wide range for $E_g$ and $T$. A and B are estimated by keeping function $f$ and $g$ as parameters while fitting the $L$ vs $S$ data with the proposed equation. As a result, apart from A and B,  we got the information of how f and g are varying as a function of $Eg$ and $T$ and from analyzing that function f and g are determined giving a complete expression of $L$ as function of $S$ with $E_g$ and $T$ as parameters. The final equation for $L$ can be expressed as:

\begin{equation} \label{new}
    L  = 1.1 + 1.35 exp \left[\frac{-|S|}{\left[162 - 128exp \left(-\frac{Eg}{0.614}\right)     \right]  \left[\frac{15.71}{\sqrt{T}} + 0.0925   \right]}       \right]
\end{equation} 
where $L$ is in 10$^{-8}$ W$\Omega$K$^{-2}$ and $S$ in $\mu$VK$^{-1}$. \\

In order to show the validity of proposed equation, we have shown a good agreement between simulated $L$ as a function of $S$ using equation~\ref{see} and~\ref{l} and using our proposed equation~\ref{new}. Figure~\ref{fit}a shows the $L$ at 300K with changing band gaps. The dot represents the data obtained by solving equation~\ref{see} and~\ref{l} whereas the line represents value of $L$ generated using equation~\ref{new}. The values are in good agreement. Now to see if the equation hold for other temperatures we have done a similar simulation for band gap of 0.2, 0.4 and 0.6 with temperature of 300 K, 600 K, and 900 K as shown in figure~\ref{fit}b,c,d. In every case our proposed equation produces value of $L$ close to the value produced by solving equation~\ref{see} and~\ref{l}. This concludes the validity of our equation which can be used instead of equation~\ref{see} and~\ref{l}.
 
As a test case, we have considered a few examples where APS is the dominant scattering mechanism. One of the best example is PbTe system as shown in the figure~\ref{example}. The $L$ calculated using equation~\ref{new} fits perfectly with the reported value of $L$ obtained using equation~\ref{see} and~\ref{l}~\cite{lalonde2011reevaluation}. Another example of Sb doped ZrNiSn shows small deviation from the reported value which was calculated by considering Acoustic phonon, Polar and Alloy scattering~\cite{xie2014intrinsic}. In this case just using the SKB model is not so accurate but still the SKB estimate better value of $L$ compared to SPB model as shown with pink (SKB) and black (SPB) lines. Our proposed equation provides an easy way to use SKB model to estimate $L$ for system with small band gaps. The systems with low band gap where complex scattering mechanism are in play, using SKB model with our proposed equation will provide more accurate value than using SPB model. 
Overall, we have proposed an equation to estimate $L$ considering non parabolicity into account which require, experimental Seebeck, band gap and measurement temperature. Using our equation one can estimate the lattice thermal conductivity more accurately.

\newpage
\begin{acknowledgments}
AK acknowledge Prime Minister Research Fellowship for financial support. AK also acknowledge Prof. Surjeet Singh and Dr. Prasenjit Ghosh for valuable discussions.
\end{acknowledgments}

\bibliography{main}

\begin{thebibliography}{14}%
\makeatletter
\providecommand \@ifxundefined [1]{%
 \@ifx{#1\undefined}
}%
\providecommand \@ifnum [1]{%
 \ifnum #1\expandafter \@firstoftwo
 \else \expandafter \@secondoftwo
 \fi
}%
\providecommand \@ifx [1]{%
 \ifx #1\expandafter \@firstoftwo
 \else \expandafter \@secondoftwo
 \fi
}%
\providecommand \natexlab [1]{#1}%
\providecommand \enquote  [1]{``#1''}%
\providecommand \bibnamefont  [1]{#1}%
\providecommand \bibfnamefont [1]{#1}%
\providecommand \citenamefont [1]{#1}%
\providecommand \href@noop [0]{\@secondoftwo}%
\providecommand \href [0]{\begingroup \@sanitize@url \@href}%
\providecommand \@href[1]{\@@startlink{#1}\@@href}%
\providecommand \@@href[1]{\endgroup#1\@@endlink}%
\providecommand \@sanitize@url [0]{\catcode `\\12\catcode `\$12\catcode `\&12\catcode `\#12\catcode `\^12\catcode `\_12\catcode `\%12\relax}%
\providecommand \@@startlink[1]{}%
\providecommand \@@endlink[0]{}%
\providecommand \url  [0]{\begingroup\@sanitize@url \@url }%
\providecommand \@url [1]{\endgroup\@href {#1}{\urlprefix }}%
\providecommand \urlprefix  [0]{URL }%
\providecommand \Eprint [0]{\href }%
\providecommand \doibase [0]{http://dx.doi.org/}%
\providecommand \selectlanguage [0]{\@gobble}%
\providecommand \bibinfo  [0]{\@secondoftwo}%
\providecommand \bibfield  [0]{\@secondoftwo}%
\providecommand \translation [1]{[#1]}%
\providecommand \BibitemOpen [0]{}%
\providecommand \bibitemStop [0]{}%
\providecommand \bibitemNoStop [0]{.\EOS\space}%
\providecommand \EOS [0]{\spacefactor3000\relax}%
\providecommand \BibitemShut  [1]{\csname bibitem#1\endcsname}%
\let\auto@bib@innerbib\@empty
\bibitem [{\citenamefont {Jones}\ and\ \citenamefont {March}(1985)}]{jones1985theoretical}%
  \BibitemOpen
  \bibfield  {author} {\bibinfo {author} {\bibfnamefont {W.}~\bibnamefont {Jones}}\ and\ \bibinfo {author} {\bibfnamefont {N.~H.}\ \bibnamefont {March}},\ }\href@noop {} {\emph {\bibinfo {title} {Theoretical solid state physics}}},\ Vol.~\bibinfo {volume} {35}\ (\bibinfo  {publisher} {Courier Corporation},\ \bibinfo {year} {1985})\BibitemShut {NoStop}%
\bibitem [{\citenamefont {Snyder}\ and\ \citenamefont {Toberer}(2008)}]{snyder2008complex}%
  \BibitemOpen
  \bibfield  {author} {\bibinfo {author} {\bibfnamefont {G.~J.}\ \bibnamefont {Snyder}}\ and\ \bibinfo {author} {\bibfnamefont {E.~S.}\ \bibnamefont {Toberer}},\ }\bibfield  {title} {\enquote {\bibinfo {title} {Complex thermoelectric materials},}\ }\href@noop {} {\bibfield  {journal} {\bibinfo  {journal} {Nature materials}\ }\textbf {\bibinfo {volume} {7}},\ \bibinfo {pages} {105--114} (\bibinfo {year} {2008})}\BibitemShut {NoStop}%
\bibitem [{\citenamefont {May}, \citenamefont {Fleurial},\ and\ \citenamefont {Snyder}(2008)}]{may2008thermoelectric}%
  \BibitemOpen
  \bibfield  {author} {\bibinfo {author} {\bibfnamefont {A.~F.}\ \bibnamefont {May}}, \bibinfo {author} {\bibfnamefont {J.-P.}\ \bibnamefont {Fleurial}}, \ and\ \bibinfo {author} {\bibfnamefont {G.~J.}\ \bibnamefont {Snyder}},\ }\bibfield  {title} {\enquote {\bibinfo {title} {Thermoelectric performance of lanthanum telluride produced via mechanical alloying},}\ }\href@noop {} {\bibfield  {journal} {\bibinfo  {journal} {Physical Review B}\ }\textbf {\bibinfo {volume} {78}},\ \bibinfo {pages} {125205} (\bibinfo {year} {2008})}\BibitemShut {NoStop}%
\bibitem [{\citenamefont {Toberer}, \citenamefont {Baranowski},\ and\ \citenamefont {Dames}(2012)}]{toberer2012advances}%
  \BibitemOpen
  \bibfield  {author} {\bibinfo {author} {\bibfnamefont {E.~S.}\ \bibnamefont {Toberer}}, \bibinfo {author} {\bibfnamefont {L.~L.}\ \bibnamefont {Baranowski}}, \ and\ \bibinfo {author} {\bibfnamefont {C.}~\bibnamefont {Dames}},\ }\bibfield  {title} {\enquote {\bibinfo {title} {Advances in thermal conductivity},}\ }\href@noop {} {\bibfield  {journal} {\bibinfo  {journal} {Annual Review of Materials Research}\ }\textbf {\bibinfo {volume} {42}},\ \bibinfo {pages} {179--209} (\bibinfo {year} {2012})}\BibitemShut {NoStop}%
\bibitem [{\citenamefont {Lukas}\ \emph {et~al.}(2012)\citenamefont {Lukas}, \citenamefont {Liu}, \citenamefont {Joshi}, \citenamefont {Zebarjadi}, \citenamefont {Dresselhaus}, \citenamefont {Ren}, \citenamefont {Chen},\ and\ \citenamefont {Opeil}}]{lukas2012experimental}%
  \BibitemOpen
  \bibfield  {author} {\bibinfo {author} {\bibfnamefont {K.~C.}\ \bibnamefont {Lukas}}, \bibinfo {author} {\bibfnamefont {W.}~\bibnamefont {Liu}}, \bibinfo {author} {\bibfnamefont {G.}~\bibnamefont {Joshi}}, \bibinfo {author} {\bibfnamefont {M.}~\bibnamefont {Zebarjadi}}, \bibinfo {author} {\bibfnamefont {M.~S.}\ \bibnamefont {Dresselhaus}}, \bibinfo {author} {\bibfnamefont {Z.}~\bibnamefont {Ren}}, \bibinfo {author} {\bibfnamefont {G.}~\bibnamefont {Chen}}, \ and\ \bibinfo {author} {\bibfnamefont {C.~P.}\ \bibnamefont {Opeil}},\ }\bibfield  {title} {\enquote {\bibinfo {title} {Experimental determination of the lorenz number in cu 0.01 bi 2 te 2.7 se 0.3 and bi 0.88 sb 0.12},}\ }\href@noop {} {\bibfield  {journal} {\bibinfo  {journal} {Physical Review B}\ }\textbf {\bibinfo {volume} {85}},\ \bibinfo {pages} {205410} (\bibinfo {year} {2012})}\BibitemShut {NoStop}%
\bibitem [{\citenamefont {Pei}\ \emph {et~al.}(2011)\citenamefont {Pei}, \citenamefont {Shi}, \citenamefont {LaLonde}, \citenamefont {Wang}, \citenamefont {Chen},\ and\ \citenamefont {Snyder}}]{pei2011convergence}%
  \BibitemOpen
  \bibfield  {author} {\bibinfo {author} {\bibfnamefont {Y.}~\bibnamefont {Pei}}, \bibinfo {author} {\bibfnamefont {X.}~\bibnamefont {Shi}}, \bibinfo {author} {\bibfnamefont {A.}~\bibnamefont {LaLonde}}, \bibinfo {author} {\bibfnamefont {H.}~\bibnamefont {Wang}}, \bibinfo {author} {\bibfnamefont {L.}~\bibnamefont {Chen}}, \ and\ \bibinfo {author} {\bibfnamefont {G.~J.}\ \bibnamefont {Snyder}},\ }\bibfield  {title} {\enquote {\bibinfo {title} {Convergence of electronic bands for high performance bulk thermoelectrics},}\ }\href@noop {} {\bibfield  {journal} {\bibinfo  {journal} {Nature}\ }\textbf {\bibinfo {volume} {473}},\ \bibinfo {pages} {66--69} (\bibinfo {year} {2011})}\BibitemShut {NoStop}%
\bibitem [{\citenamefont {Wood}(1988)}]{wood1988materials}%
  \BibitemOpen
  \bibfield  {author} {\bibinfo {author} {\bibfnamefont {C.}~\bibnamefont {Wood}},\ }\bibfield  {title} {\enquote {\bibinfo {title} {Materials for thermoelectric energy conversion},}\ }\href@noop {} {\bibfield  {journal} {\bibinfo  {journal} {Reports on progress in physics}\ }\textbf {\bibinfo {volume} {51}},\ \bibinfo {pages} {459} (\bibinfo {year} {1988})}\BibitemShut {NoStop}%
\bibitem [{\citenamefont {Kim}\ \emph {et~al.}(2015)\citenamefont {Kim}, \citenamefont {Gibbs}, \citenamefont {Tang}, \citenamefont {Wang},\ and\ \citenamefont {Snyder}}]{kim2015characterization}%
  \BibitemOpen
  \bibfield  {author} {\bibinfo {author} {\bibfnamefont {H.-S.}\ \bibnamefont {Kim}}, \bibinfo {author} {\bibfnamefont {Z.~M.}\ \bibnamefont {Gibbs}}, \bibinfo {author} {\bibfnamefont {Y.}~\bibnamefont {Tang}}, \bibinfo {author} {\bibfnamefont {H.}~\bibnamefont {Wang}}, \ and\ \bibinfo {author} {\bibfnamefont {G.~J.}\ \bibnamefont {Snyder}},\ }\bibfield  {title} {\enquote {\bibinfo {title} {Characterization of lorenz number with seebeck coefficient measurement},}\ }\href@noop {} {\bibfield  {journal} {\bibinfo  {journal} {APL materials}\ }\textbf {\bibinfo {volume} {3}} (\bibinfo {year} {2015})}\BibitemShut {NoStop}%
\bibitem [{\citenamefont {LaLonde}, \citenamefont {Pei},\ and\ \citenamefont {Snyder}(2011)}]{lalonde2011reevaluation}%
  \BibitemOpen
  \bibfield  {author} {\bibinfo {author} {\bibfnamefont {A.~D.}\ \bibnamefont {LaLonde}}, \bibinfo {author} {\bibfnamefont {Y.}~\bibnamefont {Pei}}, \ and\ \bibinfo {author} {\bibfnamefont {G.~J.}\ \bibnamefont {Snyder}},\ }\bibfield  {title} {\enquote {\bibinfo {title} {Reevaluation of pbte 1- x i x as high performance n-type thermoelectric material},}\ }\href@noop {} {\bibfield  {journal} {\bibinfo  {journal} {Energy \& Environmental Science}\ }\textbf {\bibinfo {volume} {4}},\ \bibinfo {pages} {2090--2096} (\bibinfo {year} {2011})}\BibitemShut {NoStop}%
\bibitem [{\citenamefont {Wang}\ \emph {et~al.}(2012)\citenamefont {Wang}, \citenamefont {Pei}, \citenamefont {LaLonde},\ and\ \citenamefont {Snyder}}]{wang2012weak}%
  \BibitemOpen
  \bibfield  {author} {\bibinfo {author} {\bibfnamefont {H.}~\bibnamefont {Wang}}, \bibinfo {author} {\bibfnamefont {Y.}~\bibnamefont {Pei}}, \bibinfo {author} {\bibfnamefont {A.~D.}\ \bibnamefont {LaLonde}}, \ and\ \bibinfo {author} {\bibfnamefont {G.~J.}\ \bibnamefont {Snyder}},\ }\bibfield  {title} {\enquote {\bibinfo {title} {Weak electron--phonon coupling contributing to high thermoelectric performance in n-type pbse},}\ }\href@noop {} {\bibfield  {journal} {\bibinfo  {journal} {Proceedings of the National Academy of Sciences}\ }\textbf {\bibinfo {volume} {109}},\ \bibinfo {pages} {9705--9709} (\bibinfo {year} {2012})}\BibitemShut {NoStop}%
\bibitem [{\citenamefont {Wang}\ \emph {et~al.}(2013)\citenamefont {Wang}, \citenamefont {Schechtel}, \citenamefont {Pei},\ and\ \citenamefont {Snyder}}]{wang2013high}%
  \BibitemOpen
  \bibfield  {author} {\bibinfo {author} {\bibfnamefont {H.}~\bibnamefont {Wang}}, \bibinfo {author} {\bibfnamefont {E.}~\bibnamefont {Schechtel}}, \bibinfo {author} {\bibfnamefont {Y.}~\bibnamefont {Pei}}, \ and\ \bibinfo {author} {\bibfnamefont {G.~J.}\ \bibnamefont {Snyder}},\ }\bibfield  {title} {\enquote {\bibinfo {title} {High thermoelectric efficiency of n-type pbs},}\ }\href@noop {} {\bibfield  {journal} {\bibinfo  {journal} {Advanced Energy Materials}\ }\textbf {\bibinfo {volume} {3}},\ \bibinfo {pages} {488--495} (\bibinfo {year} {2013})}\BibitemShut {NoStop}%
\bibitem [{\citenamefont {Bhandari}\ and\ \citenamefont {Rowe}(1985)}]{bhandari1985electronic}%
  \BibitemOpen
  \bibfield  {author} {\bibinfo {author} {\bibfnamefont {C.}~\bibnamefont {Bhandari}}\ and\ \bibinfo {author} {\bibfnamefont {D.}~\bibnamefont {Rowe}},\ }\bibfield  {title} {\enquote {\bibinfo {title} {Electronic contribution to the thermal conductivity of narrow band gap semiconductors-effect of non-parabolicity of bands},}\ }\href@noop {} {\bibfield  {journal} {\bibinfo  {journal} {Journal of Physics D: Applied Physics}\ }\textbf {\bibinfo {volume} {18}},\ \bibinfo {pages} {873} (\bibinfo {year} {1985})}\BibitemShut {NoStop}%
\bibitem [{\citenamefont {Ravich}(2013)}]{ravich2013semiconducting}%
  \BibitemOpen
  \bibfield  {author} {\bibinfo {author} {\bibfnamefont {I.~I.}\ \bibnamefont {Ravich}},\ }\href@noop {} {\emph {\bibinfo {title} {Semiconducting lead chalcogenides}}},\ Vol.~\bibinfo {volume} {5}\ (\bibinfo  {publisher} {Springer Science \& Business Media},\ \bibinfo {year} {2013})\BibitemShut {NoStop}%
\bibitem [{\citenamefont {Xie}\ \emph {et~al.}(2014)\citenamefont {Xie}, \citenamefont {Wang}, \citenamefont {Fu}, \citenamefont {Liu}, \citenamefont {Snyder}, \citenamefont {Zhao},\ and\ \citenamefont {Zhu}}]{xie2014intrinsic}%
  \BibitemOpen
  \bibfield  {author} {\bibinfo {author} {\bibfnamefont {H.}~\bibnamefont {Xie}}, \bibinfo {author} {\bibfnamefont {H.}~\bibnamefont {Wang}}, \bibinfo {author} {\bibfnamefont {C.}~\bibnamefont {Fu}}, \bibinfo {author} {\bibfnamefont {Y.}~\bibnamefont {Liu}}, \bibinfo {author} {\bibfnamefont {G.~J.}\ \bibnamefont {Snyder}}, \bibinfo {author} {\bibfnamefont {X.}~\bibnamefont {Zhao}}, \ and\ \bibinfo {author} {\bibfnamefont {T.}~\bibnamefont {Zhu}},\ }\bibfield  {title} {\enquote {\bibinfo {title} {The intrinsic disorder related alloy scattering in zrnisn half-heusler thermoelectric materials},}\ }\href@noop {} {\bibfield  {journal} {\bibinfo  {journal} {Scientific reports}\ }\textbf {\bibinfo {volume} {4}},\ \bibinfo {pages} {6888} (\bibinfo {year} {2014})}\BibitemShut {NoStop}%
\end{thebibliography}%

\end{document}